\documentclass{PoS}
\usepackage[dvipsnames]{xcolor}
\usepackage{heppennames2}
\usepackage{lhchiggs}
\usepackage{cernunits}
\usepackage{pifont}
\usepackage{url}
\usepackage[utf8]{inputenc}
\DeclareRobustCommand{\PA}{\HepParticle{A}{}{}\Xspace}
\DeclareRobustCommand{\PV}{\HepParticle{V}{}{}\Xspace}
\DeclareRobustCommand{\PX}{\HepParticle{X}{}{}\Xspace}

\DeclareRobustCommand{\Pf}{\HepParticle{f}{}{}\Xspace}
\DeclareRobustCommand{\PAf}{\HepAntiParticle{\Pf}{}{}\Xspace}

\DeclareRobustCommand{\PL}{\HepParticle{L}{}{}\Xspace}

\DeclareRobustCommand{\PQQ}{\HepParticle{Q}{}{}\Xspace}

\newcommand{\PHH}{\PH\PH}
\newcommand{\PAA}{\PA\PA}
\newcommand{\PAZ}{\PA\PZ}
\newcommand{\PZA}{\PZ\PA}
\newcommand{\PZZ}{\PZ\PZ}

\newcommand{\PVV}{\PV\PV}

\newcommand{\myQED}{\mathrm{QED}}

\newcommand{\ssA}{{\mathrm{A}}}

\newcommand{\ssF}{{\mathrm{F}}}

\newcommand{\ssR}{{\mathrm{R}}}

\newcommand{\ssD}{{\mathrm{D}}}

\newcommand{\ssP}{{\mathrm{P}}}

\newcommand{\ssS}{{\mathrm{S}}}

\newcommand{\ssV}{{\mathrm{V}}}

\newcommand{\ssN}{{\mathrm{N}}}

\newcommand{\mrc}{{\mathrm{c}}}

\newcommand{\mrs}{{\mathrm{s}}}
\newcommand{\mrv}{{\mathrm{v}}}

\newcommand{\ssT}{{\mathrm{T}}}

\newcommand{\bqas}{\begin{eqnarray*}}
\newcommand{\eqas}{\end{eqnarray*}}
\newcommand{\nl}{\nonumber\\}

\newcommand{\lpar}{\left(}                            
\newcommand{\rpar}{\right)}

\newcommand{\bq}{\begin{equation}}                    
\newcommand{\eq}{\end{equation}}
\newcommand{\bqa}{\arraycolsep 0.14em\begin{eqnarray}}
\newcommand{\eqa}{\end{eqnarray}}
\newcommand{\ba}[1]{\begin{array}{#1}}
\newcommand{\ea}{\end{array}}
\newcommand{\ben}{\begin{enumerate}}
\newcommand{\een}{\end{enumerate}}
\newcommand{\bei}{\begin{itemize}}
\newcommand{\eei}{\end{itemize}}

\newcommand{\sect}[1]{Section~\ref{#1}}

\newcommand{\Bref}[1]{Ref.~\cite{#1}}
\newcommand{\Brefs}[1]{Refs.~\cite{#1}}

\newcommand{\eg}{e.g.\xspace}
\newcommand{\ie}{i.e.\xspace}
\newcommand{\etc}{etc.\@\xspace}

\newcommand{\mz}{\mathswitch{M_{\PZ}}}

\newcommand{\mzc}{\mathswitch{M^3_{\PZ}}}

\newcommand{\mhs}{\mathswitch{M^2_{\sPH}}}

\newcommand{\sPL}{{\scriptscriptstyle{\PL}}}

\newcommand{\sPH}{{\scriptscriptstyle{\PH}}}

\newcommand{\sPW}{{\scriptscriptstyle{\PW}}}

\newcommand{\sPB}{{\scriptscriptstyle{\PB}}}

\newcommand{\sPHAA}{{\scriptscriptstyle{\PH\PAA}}}
\newcommand{\sPHAZ}{{\scriptscriptstyle{\PH\PA\PZ}}}

\newcommand{\sPHZZ}{{\scriptscriptstyle{\PH\PZ\PZ}}}
\newcommand{\sPHWW}{{\scriptscriptstyle{\PH\PW\PW}}}

\newcommand{\emc}{\mathswitch{\gamma}}

\newcommand{\muR}{\mathswitch{\mu_{\ssR}}}

\newcommand{\muRs}{\mathswitch{\mu^2_{\ssR}}}

\newcommand{\brem}{{\mbox{\scriptsize brem}}}

\newcommand{\tree}{{\mbox{\scriptsize tree}}}
\newcommand{\icdiv}{{\mbox{\scriptsize div}}}

\newcommand{\ren}{{\mbox{\scriptsize ren}}}

\newcommand{\ep}{\mathswitch \varepsilon}

\newcommand{\spro}[2]{{#1}\cdot{#2}}

\newcommand{\mrS}{{\mathrm{S}}}
\newcommand{\mrA}{{\mathrm{A}}}

\newcommand{\mrF}{{\mathrm{F}}}

\newcommand{\mrT}{{\mathrm{T}}}

\newcommand{\mrV}{{\mathrm{V}}}

\newcommand{\mrZ}{{\mathrm{Z}}}

\providecommand{\dnuma}{{\Large{\ding{172}}}\Xspace}
\providecommand{\dnumb}{{\Large{\ding{173}}}\Xspace}
\providecommand{\dnumc}{{\Large{\ding{174}}}\Xspace}

\newtheorem{theorem}{Theorem}[section]

\newtheorem{proposition}[theorem]{Proposition}

\newenvironment{remark}[1][Remark]{\begin{trivlist}
\item[\hskip \labelsep {\bfseries #1}]}{\end{trivlist}}

\newcommand{\Ope}{{\cal O}}

\newcommand{\OpWs}{\Ope^{(6)}_{\upphi\,\scriptscriptstyle{\PW}}}
\newcommand{\OpWe}{\Ope^{(8)}_{\upphi\,\scriptscriptstyle{\PW}}}
\newcommand{\myGF}{G_{\mrF}}
\DeclareRobustCommand{\PK}{\HepParticle{\upPhi}{}{}\Xspace}

\DeclareRobustCommand{\PKdag}{\HepParticle{\PK}{}{\dagger}\Xspace}
\newcommand{\KdK}{\lpar \PKdag\,\PK\rpar }

\DeclareRobustCommand{\PAf}{\HepAntiParticle{\Pf}{}{}\Xspace}
\newcommand{\stw}{\mrs_{_{\theta}}}             
\newcommand{\ctw}{\mrc_{_{\theta}}}
\newcommand{\stws}{\mrs_{_{\theta}}^2}
\newcommand{\ctws}{\mrc_{_{\theta}}^2}
\newcommand{\ctwc}{\mrc_{_{\theta}}^3}

\DeclareRobustCommand{\PWp}{\HepParticle{\PW}{}{+}\Xspace}
\DeclareRobustCommand{\PWm}{\HepParticle{\PW}{}{-}\Xspace}

\newcommand{\app}{{\mbox{\scriptsize app}}}

\newcommand{\apLA}{a_{\upphi\,\sPL\,\scriptscriptstyle{\PA}}}

\newcommand{\aQW}{a_{\scriptscriptstyle{\PQQ\PW}}}
\newcommand{\aAA}{a_{\scriptscriptstyle{\PAA}}}
\newcommand{\aZZ}{a_{\scriptscriptstyle{\PZZ}}}
\newcommand{\aAZ}{a_{\scriptscriptstyle{\PAZ}}}

\newcommand{\apBox}{a_{\upphi\,\scriptscriptstyle{\Box}}}

\newcommand{\apD}{a_{\upphi\,\scriptscriptstyle{\PD}}}

\newcommand{\adp}{a_{\PQd\,\upphi}}

\newcommand{\auW}{a_{\PQu\,\scriptscriptstyle{\PW}}}

\newcommand{\auB}{a_{\PQu\,\scriptscriptstyle{\PB}}}

\newcommand{\auWB}{a_{\PQu\,\scriptscriptstyle{\PW\PB}}}
\newcommand{\auBW}{a_{\PQu\,\scriptscriptstyle{\PB\PW}}}

\newcommand{\adW}{a_{\PQd\,\scriptscriptstyle{\PW}}}

\newcommand{\adB}{a_{\PQd\,\scriptscriptstyle{\PB}}}
\newcommand{\adWB}{a_{\PQd\,\scriptscriptstyle{\PW\PB}}}
\newcommand{\adBW}{a_{\PQd\,\scriptscriptstyle{\PB\PW}}}

\newcommand{\alW}{a_{\Pl\,\scriptscriptstyle{\PW}}}
\newcommand{\alB}{a_{\Pl\,\scriptscriptstyle{\PB}}}
\newcommand{\alWB}{a_{\Pl\,\scriptscriptstyle{\PW\PB}}}
\newcommand{\alBW}{a_{\Pl\,\scriptscriptstyle{\PB\PW}}}
\newcommand{\aplV}{a_{\upphi\,\Pl\,\scriptscriptstyle{\PV}}}
\newcommand{\aplA}{a_{\upphi\,\Pl\,\scriptscriptstyle{\PA}}}

\newcommand{\apuV}{a_{\upphi\,\PQu\,\scriptscriptstyle{\PV}}}

\newcommand{\apuA}{a_{\upphi\,\PQu\,\scriptscriptstyle{\PA}}}

\newcommand{\apdV}{a_{\upphi\,\PQd\,\scriptscriptstyle{\PV}}}

\newcommand{\apdA}{a_{\upphi\,\PQd\,\scriptscriptstyle{\PA}}}

\newcommand{\apn}{a_{\upphi\,\PGn}}
\newcommand{\aplo}{a^{(1)}_{\upphi\,\Pl}}
\newcommand{\aplt}{a^{(3)}_{\upphi\,\Pl}}
\newcommand{\apqo}{a^{(1)}_{\upphi\,\PQq}}

\newcommand{\apqt}{a^{(3)}_{\upphi\,\PQq}}

\newcommand{\apu}{a_{\upphi\,\PQu}}
\newcommand{\apd}{a_{\upphi\,\PQd}}
\newcommand{\apl}{a_{\upphi\,\Pl}}

\newcommand{\vle}{\mathrm{v}_{\Pl}}

\newcommand{\sla}[1]{/\!\!\!\!\!#1}
\newcommand{\DUV}{\Delta_{\mathrm{UV}}}

\newcommand{\mcA}{\mathcal{A}}

\newcommand{\mcD}{\mathcal{D}}

\newcommand{\mcO}{\mathcal{O}}
\newcommand{\mcP}{\mathcal{P}}

\newcommand{\mcF}{\mathcal{F}}

\newcommand{\mcT}{\mathcal{T}}

\newcommand{\mrdim}{\mathrm{dim}}
\newcommand{\cdim}{\mathrm{codim}}
\newcommand{\spc}{\,,}
\newcommand{\spp}{\,\,.}

\newcommand{\redeq}[1]{{\color{BrickRed}{#1}}}
\usepackage{heppennames2}
\usepackage{lhchiggs}
\usepackage{cernunits}
\allowdisplaybreaks
\title{NLO Standard model effective field theory for Higgs and EW precision data}

\ShortTitle{SMEFT}

\author{\speaker{Giampiero Passarino}\thanks{Work supported by the Research Executive 
Agency (REA) of the European Union under the Grant Agreement PITN-GA-2012-316704 (HiggsTools).}
\\
Dipartimento di Fisica Teorica, Universit\`a di Torino, 
INFN, Sezione di Torino, Italy\\
E-mail: \email{giampiero@to.infn.it}}

\abstract{A set of constructs, definitions, and propositions that present
a systematic view of the Standard Model Effective Field Theory (SMEFT), \ie how the influence 
of higher energy processes is localizable in a few structural properties which can be captured 
by a handful of Wilson coefficients.}

\FullConference{Loops and Legs in Quantum Field Theory\\
		 24 April - 26 April 2016\\
		 Leipzig, Germany}

\begin{document}
\section{Introduction}
In these proceedings we provide a consistency proof (not only power counting, but a proof 
that proves that there are enough Wilson coefficients) of quasi{-}renormalizability in SMEFT. 
Theory deals with the well founded theoretical results obtained from first principles, while 
phenomenology deals with not so well founded effective models with a smaller domain of 
application. For a definition see \Bref{Hartmann2001}.

Mathematics suffers from some of the same inherent difficulties as theoretical physics: 
great successes during the $2$0th century, increasing difficulties to do better, as the easier 
problems get solved. The lesson of experiments $1973$ - today: it is extremely difficult to 
find a flaw in the Standard Model (SM): maybe the SM includes elements of a truly fundamental 
theory. But then how can one hope to make progress without experimental guidance? One should 
pay close attention to what we do not understand precisely about the SM even if the standard 
prejudice is ``that's a hard technical problem, and solving it won't change anything''.

There is a conventional vision: some very different physics occurs at Planck scale, SM is just 
an effective field theory. What about the next SM? A new weakly coupled renormalizable model? 
A tower of EFTs? A different vision: is the SM close to a fundamental theory?

It is possible that at some very large energy scale, all nonrenormalizable interactions 
disappear. This seems unlikely, given the difficulty with gravity. It is possible 
that the rules change drastically, it may even be possible that there is no end, simply 
more and more scales. 
This prompts the important question whether there is a last fundamental theory in this tower 
of EFTs which supersede each other with rising energies. Some people conjecture that this 
deeper theory could be a string theory, \ie a theory which is not a field theory any more. 
Or should one ultimately expect from physics theories that they are only valid as approximations 
and in a limited domain~\cite{Hartmann2001,Castellani}?
Alternatively, one should not resort to arguments involving gravity: let us banish further 
thoughts about gravity and the damage it could do to the weak scale~\cite{Wells:2016luz}.

When looking for ultraviolet (UV) completions of the SM the following remarks are relevant:
there are $45$ spin $1/2$ and $27$ spin $1$ dof, only one spin $0$? If there are more the
present knowledge requires a hierarchy of VEVs which, once again, is a serious fine-tuning
problem. Why are all mixings small? Is it accidental or systematic (\ie a new symmetry)?
The real problem when dealing with UV completions is that one model is falsifiable, but an endless 
stream of them is not.
\section{Theoretical framework}
Back to the ``more and more scales'' scenario. Let's undergo revision (SMEFT) but it is an error 
to believe that rigour is the enemy of simplicity. On the contrary we find it confirmed by 
numerous examples that the rigorous method is at the same time the simpler and the more easily 
comprehended. 
To summarize: there is a need for a consistent theoretical framework in which deviations from 
the SM (or NextSM) predictions can be calculated, every $20$ bogus hypotheses you test, one of 
them will give you a $p$ of $< 0.05$. Such a framework should be applicable to comprehensively 
describe measurements in all sectors of particle physics: LHC Higgs measurements, past EWPD, \etc
Consider the SM augmented with the inclusion of higher dimensional operators and call it 
$\mrT_1$; it is not strictly renormalizable. Although workable to all orders, $\mrT_1$ 
fails above a certain scale, $\Lambda_1$.
Consider any BSM model that is strictly renormalizable and respects unitarity ($\mrT_2$); its 
parameters can be fixed by comparison with data, while masses of heavy states are presently 
unknown. Note that $\mrT_1 \not= \mrT_2$ in the UV but must have the same IR behavior.
Consider now the whole set of data below $\Lambda_1$: $\mrT_1$ should be able to explain them 
by fitting Wilson coefficients, $\mrT_2$ adjusting the masses of heavy states (as SM did with 
the Higgs mass at LEP) should be able to explain the data. 
Goodness of both explanations are crucial in understanding how well they match and how 
reasonable is to use $\mrT_1$ instead of the full $\mrT_2$. Does $\mrT_2$ explain everything? 
Certainly not, but it should be able to explain something more than $\mrT_1$. 
We could now define $\mrT_3$ as $\mrT_2$ augmented with (its own) 
higher dimensional operators; it is valid up to a scale $\Lambda_2$. Etc.
\subsection{SMEFT}
The construction of the SMEFT, to all orders, is not based on assumptions on 
the size of the Wilson coefficients of the higher dimensional operators.
Restricting to a particular UV case is not an integral part of a general SMEFT 
treatment and various cases can be chosen once the general calculation is performed. 
If the value of Wilson coefficients in broad UV scenarios could be inferred in 
general this would be of significant scientific value.

{\underline{To summarize}}: constructing SMEFT is based on the fact that experiments occur 
at finite energy and ``measure'' an effective action $\mrS^{\mathrm{eff}}(\Lambda)$;
whatever QFT should give low energy $\mrS^{\mathrm{eff}}(\Lambda)\,,\;\forall\,\Lambda < \infty$.
One also assumes that there is no fundamental scale above which $\mrS^{\mathrm{eff}(\Lambda)}$ 
is not defined~\cite{Costello2011} and $\mrS^{\mathrm{eff}}(\Lambda)$ loses its predictive power 
if a process at $E = \Lambda$ requires $\infty$ renormalized parameters~\cite{Preskill:1990fr}.
It is remarkable that when constructive proofs are provided, their simplicity always seems 
to detract from their originality.
\subsection{The UV connection}
The SMEFT approach is based on the following 
Lagrangian~\cite{Passarino:2012cb,Ghezzi:2015vva,NLOnote,Hartmann:2015oia,Hartmann:2015aia}:
\bq
\mcA = \sum_{n=\ssN}^{\infty}\,g^n\,\mcA^{(4)}_n +
       \sum_{n=\ssN_6}^{\infty}\,\sum_{l=1}^n\,\sum_{k=1}^{\infty}\,
        g^n\,g^l_{4+2\,k}\,
        \mcA^{(4+2\,k)}_{n\,l\,k} \spc
\label{SMEFTLag}
\eq
where we use the ``Warsaw'' basis~\cite{Grzadkowski:2010es}. Here $g$ is the $SU(2)$ 
coupling constant and 
\bq
g_{4+2\,k} = 1/(\sqrt{2}\,G_{\ssF}\,\Lambda^2)^k = g^k_6 \spc
\eq
$G_{\ssF}$ is the Fermi coupling constant and $\Lambda$ is the scale around which
new physics (NP) must be resolved.
For each process $N$ defines the $\mrdim = 4$ leading order (LO) (\eg $N = 1$ for $\PH \to \PV\PV$
\etc but $N = 3$ for $\PH \to \PGg\PGg$). $N_6 = N$ for tree initiated processes and
$N - 2$ for loop initiated ones. Single insertions of $\mrdim = 6$ operators defines 
next-to-leading (NLO) SMEFT.
Ex: $\PH\PGg\PGg$ (tree) vertex generated by $\OpWs = \KdK\,\ssF^{a\,\mu\nu}\,\ssF^a_{\mu\nu}$,
by $\OpWe= \PKdag\,\ssF^{a\,\mu\nu}\,\ssF^a_{\mu\rho}\,\ssD^{\rho}\,\ssD_{\nu}\,\PK$ \etc

A simple SMEFT ordertable for tree initiated $1 \to 2$ processes is as follows
(N.B. $g_8$ denotes a single $\Ope^{(8)}$ insertion, $g^2_6$ denotes two, distinct, 
$\Ope^{(6)}$ insertions):
\[
\begin{array}{llll}
g\,/\,\mrdim & \longrightarrow & & \\
\downarrow   & g\,\mcA^{(4)}_1  &
               \; + \; g\,g_6\,\mcA^{(6)}_{1,1,1} &
               \; + \; g\,g_8\,\mcA^{(8)}_{1,1,2} \\
             & g^3\,\mcA^{(4)}_3  &
               \; + \; g^3\,g_6\,\mcA^{(6)}_{3,1,1} &
               \; + \; g^3\,g^2_6\,\mcA^{(6)}_{3,2,1} \\
             &  \dots\dots & \dots\dots & \dots\dots \\
\end{array}
\]
\bei

\item[\dnuma] $g\,g_6\,\mcA^{(6)}_{1,1,1}$ defines LO SMEFT. There is also RG-improved LO 
              and missing higher orders uncertainty (MHOU) for LO SMEFT; 

\item[\dnumb] $g^3\,g_6\,\mcA^{(6)}_{3,1,1}$ defines NLO SMEFT;

\item[\dnumc] $g\,g_8\,\mcA^{(8)}_{1,1,2}$, $g^3\,g^2_6\,\mcA^{(6)}_{3,2,1}$ give MHOU for 
              NLO SMEFT.

\eei
The interplay between integrating out heavy scalars and the SM decoupling limit has been 
discussed in \Bref{Boggia:2016asg}. In the very general case the SM decoupling limit 
cannot be obtained by making only assumptions about one parameter.

Working in a spontaneously broken gauge theories has consequences related to the duality
$\PH{-}$VEV. We recall the concept of (naive) power counting (for a general formulation of 
power counting see \Bref{Gavela:2016bzc}): any local operator in the Lagrangian is schematically 
of the form
\bqa
{}&{}& \Ope = 
\Lambda^{-n}\,
\overbrace{
M^l\,
\underbrace{\partial^c\,
\overbrace{
{\overline\psi}^a\,\psi^b\,\lpar \Phi^{\dagger} \rpar^d\,\Phi^e\,\PA^f
          }_{\cdim}
          }^{\mathrm{N}_{\mathrm{F}}}
          }^{\mrdim}
\nl
{}&{}&
\frac{3}{2}\,(a + b) + c + d + e + f + l + n = 4 \spp
\eqa
where Lorentz, flavor and group indices have been suppressed, $\psi$ stands for a generic 
fermion fields, $\Phi$ for a generic scalar and A for a generic gauge field.
All light masses are scaled in units of the (bare) $\PW$ mass $M$. We define dimensions according 
to
\bq
\cdim\,\Ope = \frac{3}{2}\,(a + b) + c + d + e + f \spc
\qquad
\mrdim\,\Ope = \cdim + l \spp
\label{dimdef}
\eq
One loop renormalization is controlled by:
$\mrdim = 6 \spc\; \cdim = 4 \spc \; \mathrm{N}_{\mathrm{F}} > 2$. 
The hearth of the problem: a large number of operators implodes into a small number of 
coefficients, \eg there are $92$ SM vertices, $28$ CP even operators ($1$ flavor, 
$\mathrm{N}_{\psi} = 0,2$).

Debate topic for SMEFT is the choice of a ``basis'' for $\mrdim = 6$ operators. Clearly all
bases are equivalent as long as they are a ``basis'', containing the minimal set of
operators after the use of equations of motion~\cite{Grzadkowski:2010es} and respecting the 
$SU(3)\,\times\,SU(2)\,\times\,U(1)$ gauge invariance. From a more formal point of view a basis is
characterized by its closure with respect to renormalization. Equivalence of bases should always be
understood as a statement for the $\mrS\,$-matrix and not for the Lagrangian, as dictated by
the equivalence theorem, see \Brefs{Kallosh:1972ap,Arzt:1993gz}. Any phenomenological
approach that misses one of these ingredients is still acceptable for a preliminar analysis,
as long as it does not pretend to be an EFT.
Strictly speaking we are considering here the virtual part of SMEFT; of course, the real 
(emission) part of SMEFT should be included, see \sect{real}. 
\subsection{Self energies}
Our first step deals with renormalization of self-energies: here 
$\DUV = \frac{2}{4 - n} - \emc - \ln \pi - \ln\frac{\muRs}{\mu^2}$,
$n$ is space-time dimension, the loop measure is $\mu^{4 - n}\,d^nq$ and
$\muR$ is the renormalization scale.
\bqa
\ssS_{\PHH} &=& \frac{g^2}{16\,\pi^2}\,\Sigma_{\PHH} =
\frac{g^2}{16\,\pi^2}\,\lpar \Sigma^{(4)}_{\PHH} + g_6\,\Sigma^{(6)}_{\PHH} \rpar \spc
\nl
S^{\mu\nu}_{\PAA} &=& \frac{g^2}{16\,\pi^2}\,\Sigma^{\mu\nu}_{\PAA}
\qquad
\Sigma^{\mu\nu}_{\PAA} = \Pi_{\PAA}\,\ssT^{\mu\nu} \spc
\nl
S^{\mu\nu}_{\PVV} &=& \frac{g^2}{16\,\pi^2}\,\Sigma^{\mu\nu}_{\PVV} \spc
\qquad
\Sigma^{\mu\nu}_{\PVV} = \ssD_{\PVV}\,\delta^{\mu\nu} + \ssP_{\PVV}\,p^{\mu}\,p^{\nu} \spc
\nl
\ssD_{\PVV} &=& \ssD^{(4)}_{\PVV} + g_6\,\ssD^{(6)}_{\PVV} \spc
\qquad
\ssP_{\PVV} = \ssP^{(4)}_{\PVV} + g_6\,\ssP^{(6)}_{\PVV} 
\nl
S^{\mu\nu}_{\PZA} &=& \frac{g^2}{16\,\pi^2}\,\Sigma^{\mu\nu}_{\PZA} +
g_6\,\ssT^{\mu\nu}\,\aAZ \spc
\qquad
\Sigma^{\mu\nu}_{\PZA} = \Pi_{\PZA}\,\ssT^{\mu\nu} + \ssP_{\PZA}\,p^{\mu}\,p^{\nu} \spc
\nl
\ssS_{\Pf} &=& \frac{g^2}{16\,\pi^2}\,\Bigl[ \Delta_{\Pf} +
\lpar \ssV_{\Pf} - \ssA_{\Pf}\,\gamma^5 \rpar\, i \sla{p} \Bigr] \spp
\eqa
We introduce counterterms:
\bq
\mrZ_i = 1 + \frac{g^2}{16\,\pi^2}\,\lpar d\mrZ^{(4)}_i + g_6\,d\mrZ^{(6)}_i \rpar\,\DUV \spp
\eq 
With field/parameter counterterms we can make 
$\ssS_{\PHH}, \Pi_{\PAA}, \ssD_{\PVV}, \Pi_{\PZA}$, $\ssV_{\Pf}, \ssA_{\Pf}$ 
and the corresponding Dyson resummed propagators
UV finite at $\mcO(g^2\,g_6)$ , which is enough when working under the assumption that 
gauge bosons couple to conserved currents. A gauge-invariant description turns out to 
be mandatory.
\subsection{More legs}
However, field/parameter counterterms are not enough to make UV finite the Green's functions 
with more than two legs. A mixing matrix among Wilson coefficients is needed:
\bq
a_i = \sum_j\,\mrZ^{\sPW}_{ij}\,a^{\ren}_j \quad 
\mrZ^{\sPW}_{ij} = \delta_{ij} + \frac{g^2}{16\,\pi^2}\,d\mrZ^{\sPW}_{ij}\,\DUV \spp
\eq
Define the following combinations of Wilson coefficients (where 
$\stw(\ctw)$ denotes the sine(cosine) of the renormalized weak-mixing angle):
\bqa
\aZZ &=& \stws\,a_{\upphi\,\sPB} + \ctws\,a_{\upphi\,\sPW} - \stw\,\ctw\,a_{\upphi\,\sPW\sPB} \spc
\nl
\aAA &=& \ctws\,a_{\upphi\,\sPB} + \stws\,a_{\upphi\,\sPW} + \stw\,\ctw\,a_{\upphi\,\sPW\sPB} \spc
\nl
\aAZ &=& 2\,\ctw\,\stw\,\lpar a_{\upphi\,\sPW} - a_{\upphi\,\sPB} \rpar + 
             \lpar 2\,\ctws - 1 \rpar\,a_{\upphi\,\sPW\sPB} \spc
\eqa
and compute the (on-shell) decay $\underline{\PH(P) \to \PA_{\mu}(p_1)\PA_{\nu}(p_2)}$ 
where the amplitude is 
\bq
\mrA^{\mu\nu}_{\sPHAA} = \mcT_{\sPHAA}\,T^{\mu\nu} \spc
\quad
\mhs\,T^{\mu\nu} = p^{\mu}_2\,p^{\nu}_1 - \spro{p_1}{p_2}\,\delta^{\mu\nu} \spp
\eq
\begin{remark}[]
This amplitude is made UV finite by mixing $\aAA$ with $\aAA, \aAZ, \aZZ$ and $\aQW$ 
\end{remark}
Compute the (on-shell) decay $\underline{\PH(P) \to \PA_{\mu}(p_1)\PZ_{\nu}(p_2)}$.
After adding $1$PI and $1$PR components we obtain
\bq
\mrA^{\mu\nu}_{\sPHAZ} = \mcT_{\sPHAZ}\,T^{\mu\nu} 
\quad
\mhs\,T^{\mu\nu} = p^{\mu}_2\,p^{\nu}_1 - \spro{p_1}{p_2}\,\delta^{\mu\nu} 
\eq
\begin{remark}[]
This amplitude is made UV finite by mixing $\aAZ$ with $\aAA, \aAZ, \aZZ$ and $\aQW$. 
\end{remark}
Compute the (on-shell) decay $\underline{\PH(P) \to \PZ_{\mu}(p_1)\PZ_{\nu}(p_2)}$.
How to use it has been explained in \Bref{David:2015waa}. The amplitude contains 
a $\mcD_{\sPHZZ}$ part proportional to $\delta^{\mu\nu}$ and 
a $\mcP_{\sPHZZ}$ part proportional to $p^{\mu}_2\,p^{\nu}_1$. 
\begin{remark}
Mixing of $\aZZ$ with other Wilson coefficients makes $\mcP_{\sPHZZ}$ 
UV finite, while the mixing of $\apBox$ makes $\mcD_{\sPHZZ}$ UV finite.
\end{remark}
Compute the (on-shell) decay $\underline{\PH(P) \to \PWm_{\mu}(p_1)\PWp_{\nu}(p_2)}$.
This process follows the same decomposition of $\PH \to \PZ\PZ$ and it is UV finite in the
$\mrdim = 4$ part. However, for the $\mrdim = 6$ one, there are no Wilson coefficients
left free in $\mcP_{\sPHWW}$ so that its UV finiteness follows from gauge cancellations 
($\PH \to \PA\PA,\,\PA\PZ,\,\PZ\PZ,\,\PW\PW = 6\,$ Lorentz structures controlled 
by $5$ coefficients).
\begin{proposition}
This is the first part in proving closure of NLO SMEFT under renormalization. 
\end{proposition}
\begin{remark}
Mixing of $\apD$ makes $\mcD_{\sPHWW}$ UV finite. 
\end{remark}

\begin{remark}
Compute the (on-shell) decay $\underline{\PH(P) \to \PQb(p_1)\PAQb(p_2)}$. 
It is $\mrdim = 4$ UV finite and mixing of $\adp$ makes it UV finite also at $\mrdim= 6$. 
\end{remark}

\begin{remark}
Compute the (on-shell) decay $\underline{\PZ(P) \to \Pf(p_1)\PAf(p_2)}$. It is 
$\mrdim = 4$ UV finite and we introduce
\bqa
\alW &=& \stw\,\alWB + \ctw\,\alBW  \qquad \alB = \stw\,\alBW - \ctw\,\alWB \spc
\nl
\adW &=& \stw\,\adWB + \ctw\,\adBW  \qquad \adB = \stw\,\adBW - \ctw\,\adWB \spc
\nl
\auW &=& \stw\,\auWB + \ctw\,\auBW  \qquad \auB= \ctw\,\auWB - \stw\,\auBW  \spc
\eqa
\bqa
\aplt - \aplo &=&   \frac{1}{2}\,(\aplV + \aplA)  \spc
\qquad
\apl = \frac{1}{2}\,(\aplA - \aplV)  \spc
\nl
\apuV &=& \apqt + \apu + \apqo  \qquad \apuA = \apqt - \apu + \apqo  \spc
\nl
\apdV &=& \apqt - \apd - \apqo  \qquad \apdA = \apqt + \apd - \apqo  \spc
\eqa
and obtain that 
\bei
\item[$\PZ \to \PAl\Pl$] requires mixing of $\alBW, \aplA$ and $\aplV$ with other coefficients,
\item[$\PZ \to \PAQu\PQu$] requires mixing of $\auBW, \apuA$ and $\apuV$ with other coefficients,
\item[$\PZ \to \PAQd\PQd$] requires mixing of $\adBW, \apdA$ and $\apdV$ with other coefficients,
\item[$\PZ \to \PAGn\PGn$] requires mixing of $\apn= 2\,(\aplo + \aplt)$ with other coefficients.
\eei
\end{remark}
At this point we are left with the universality of the electric charge. In QED there is a
Ward identity telling us that $e$ is renormalized in terms of vacuum polarization and
Ward-Slavnov-Taylor (WST) identities allow us to generalize the argument to the full SM.
We can give a quantitative meaning to the the previous statement by saying that
the contribution from vertices (at zero momentum transfer) cancels those from
(fermion) wave function renormalization factors. Therefore, compute the vertex $\PA\PAf\Pf$ 
(at $q^2 = 0$) and the $\Pf$ wave function factor in SMEFT, proving that the WST identities can 
be extended to $\mrdim = 6$; this is non trivial since there are no free Wilson coefficients 
in these terms (after the previous steps); the (non-trivial) finiteness of $\Pep\Pem \to \PAf\Pf$ 
follows.
\begin{proposition}
This is the second part in proving closure of NLO SMEFT under renormalization. 
\end{proposition}
\subsection{The IR connection \label{real}}
Consider the decay $\PZ \to \PAl\Pl$, where the amplitude is
\bq
\mcA^{\tree}_{\mu} = g\,\mcA^{(4)}_{1\,\mu} + g\,g_6\,\mcA^{(6)}_{1\,\mu}  \spc
\eq
\bqa
\mcA^{(4)}_{1\,\mu} = \frac{1}{4\,\ctw}\,\gamma_{\mu}\,\lpar \vle + \gamma^5 \rpar  \spc
&\qquad&
\mcA^{(6)}_{1\,\mu}= \frac{1}{4}\,\gamma_{\mu}\,\lpar \mrV_{\Pl} + \mrA_{\Pl}\,\gamma^5 \rpar \spc
\eqa
\bqa
\mrV_{\Pl} &=& 
\frac{\stws}{\ctw}\,\lpar 4\,\stws - 7 \rpar\,\aAA +
\ctw\,\lpar 1 + 4\,\stws\rpar\,\aZZ +
\stw\,\lpar 4\,\stws - 3\rpar\,\aAZ 
\nl
{}&+& \frac{1}{4\,\ctw}\,\lpar 7 - \stws\rpar\,\apD +
\frac{2}{\ctw}\,\aplV \spc
\nl
\mrA_{\Pl} &=& 
\frac{\stws}{\ctw}\,\aAA + \ctw\,\aZZ + \stw\,\aAZ - \frac{1}{4\,\ctw}\,\apD + 
\frac{2}{\ctw}\,\apLA \spp
\eqa
After UV renormalization, \ie after counterterms and mixing have been introduced, we
perform analytic continuation in $n$ (space-time dimension), $n = 4 + \ep$ with
$\ep$ positive.
\begin{proposition}
The infrared/collinear part of the one-loop virtual corrections shows double factorization.
\end{proposition}
\bqa
\Gamma\lpar \PZ \to \PAl + \Pl \rpar\mid_{\icdiv} &=&        
- \frac{g^4}{384\,\pi^3}\,\mz\,\stws\,\mcF^{\virt}\,
\Bigl[ \redeq{\Gamma^{(4)}_0}\,\lpar 1 + g_6\,\Delta\Gamma \rpar + 
g_6\,\redeq{\Gamma^{(6)}_0} \Bigr] \spp
\eqa
\begin{proposition}
The infrared/collinear part of the real corrections shows double factorization. 
\end{proposition}
\bqa
\Gamma^{\app}\lpar \PZ \to \PAl + \Pl + (\PGg) \rpar &=&        
 \frac{g^4}{384\,\pi^3}\,\mz\,\stws\,\mcF^{\brem}\,
\Bigl[ \redeq{\Gamma^{(4)}_0}\,\lpar 1 + g_6\,\Delta\Gamma \rpar + 
g_6\,\redeq{\Gamma^{(6)}_0} \Bigr] \spp
\eqa
\begin{proposition}
The total = virtual ${+}$ real is IR/collinear finite at $\mcO(g^4\,g_6)$.
\end{proposition}
Assembling everything gives (terms in red give the SM answer)
\bqa
\Gamma^{\Pl}_{\myQED} &=& \redeq{\frac{3}{4}\,\Gamma^{\Pl}_0\,\frac{\alpha}{\pi}}\,\lpar
    1 + g_6\,\Delta^{(6)}_{\myQED} \rpar \spc
\quad
\Gamma^{\Pl}_0 = \frac{\myGF\,\mzc}{24\,\sqrt{2}\,\pi}\,\lpar v^2_{\Pl} + 1 \rpar
\nl
\Delta^{(6)}_{\myQED} &=& 
 2\,\lpar 2 - \stws \rpar\,\aAA +
 2\,\stws\,\aZZ +
 2\,\lpar \frac{\ctwc}{\stw} + \frac{512}{26}\,\frac{\mrv_{\PL}}{v^2_{\PL} + 1} \rpar\,\aAZ 
\nl
{}&-& \frac{1}{2}\,\frac{\ctws}{\stws}\,\apD +
 \frac{1}{v^2_{\PL} + 1}\,\delta^{(6)}_{\myQED} \spc
\nl
\delta^{(6)}_{\myQED} &=&
 \lpar 1 - 6\,\mrv_{\Pl} - \mrv_{\Pl}^2 \rpar\,\frac{1}{\ctws}\,
       \lpar \stw\,\aAA - \frac{1}{4}\,\apD \rpar 
\nl
{}&+&
 \lpar 1 + 2\,\mrv_{\Pl} - \mrv_{\Pl}^2 \rpar\,
       \lpar \aZZ + \frac{\stw}{\ctw}\,\aAZ \rpar + 
       \frac{2}{\ctws}\,\lpar \aplA + \mrv_{\Pl}\,\aplV \rpar
\eqa
\subsection{Next steps}
The $\PW\,$-decay series is almost completed; next, inclusion of triple/quadrupole gauge 
couplings, last stop before renormalizability? This brings us to gauge anomalies and anomaly 
cancellation; d'Hoker{-}Farhi~\cite{D'Hoker:1984ph}, (Wess{-}Zumino~\cite{Wess:1971yu}) terms
required? Extra symmetry? Severe problems are expected; perhaps, a deeper understanding of SMEFT, 
a low-energy limit of an underlying anomaly-free theory?
\begin{proposition}
SMEFT anomalies are {\underline{UV finite}} (it is good for renormalizability),
restoring gauge invariance order-by-order by adding finite counterterms, \ie
it is possible to quantize an anomalous theory in a manner that respects 
WSTI~\cite{Preskill:1990fr} and {\underline{local}}. The latter is good for unitarity, 
another tiny step forward.
\end{proposition}
\section{Conclusions}
NLO results have already had an important impact on the SMEFT physics program. 
LEP constraints should not be interpreted to mean that effective SMEFT parameters should be 
set to zero in LHC analyses.
It is important to preserve the original data, not just the interpretation results, as the 
estimate of the missing higher order terms can change over time, modifying the lessons drawn 
from the data and projected into the SMEFT. 
The assignment of a theoretical error for SMEFT analyses is always important. 
Considering projections for the precision to be reached in LHC RunII analyses, 
LO results for interpretations of the data in the SMEFT are challenged by consistency concerns 
and are not sufficient, if the cut off scale is in the few \UTeV range.
If the scale is below experimental sensitivity we are in trouble, but let's push constraints 
to the experimental limit consistently.
Unfortunately, ideas that require people to reorganize their picture of the world provoke 
hostility.

To conclude, the journey to the next (and next-to-next) SM may require crossing narrow straits 
of precision physics. If that is what nature has in store for us, we must equip ourselves with 
both a range of concrete models as well as a general theories.
However, each paradigm will be shown to satisfy more or less the criteria that it dictates for 
itself and to fall short of a few of those dictated by its opponent.
\bibliographystyle{elsarticle-num}
\bibliography{LL16p}{}

\begin{thebibliography}{10}
\expandafter\ifx\csname url\endcsname\relax
  \def\url#1{\texttt{#1}}\fi
\expandafter\ifx\csname urlprefix\endcsname\relax\def\urlprefix{URL }\fi
\expandafter\ifx\csname href\endcsname\relax
  \def\href#1#2{#2} \def\path#1{#1}\fi

\bibitem{Hartmann2001}
S.~Hartmann, {{Effective field theories, reduction and scientific
  explanation}}, {Studies in History and Philosophy of Modern Physics $32$B,
  $267{-}304$} (2001).

\bibitem{Castellani}
E.~Castellani, {{Reductionism, emergence, and effective field theories}},
  {Stud.~Hist.~Philos.~Mod.~Phys. $33 (2002) 251$} (2002).

\bibitem{Wells:2016luz}
J.~D. Wells,
  \href{http://inspirehep.net/record/1430916/files/arXiv:1603.06131.pdf}{{Higgs
  naturalness and the scalar boson proliferation instability problem}}, 2016.
\newblock \href {http://arxiv.org/abs/1603.06131} {\path{arXiv:1603.06131}},
  \href {http://dx.doi.org/10.1007/s11229-014-0618-8}
  {\path{doi:10.1007/s11229-014-0618-8}}.
\newline\urlprefix\url{http://inspirehep.net/record/1430916/files/arXiv:1603.0%
6131.pdf}

\bibitem{Costello2011}
K.~Costello, {{Renormalization and Effective Field Theory}}, {Mathematical
  Surveys and Monographs Volume $170$, American Mathematical Society} (2011).

\bibitem{Preskill:1990fr}
J.~Preskill, {Gauge anomalies in an effective field theory}, Annals Phys. 210
  (1991) 323--379.
\newblock \href {http://dx.doi.org/10.1016/0003-4916(91)90046-B}
  {\path{doi:10.1016/0003-4916(91)90046-B}}.

\bibitem{Passarino:2012cb}
G.~Passarino, {NLO Inspired Effective Lagrangians for Higgs Physics}, Nucl.
  Phys. B868 (2013) 416--458.
\newblock \href {http://arxiv.org/abs/1209.5538} {\path{arXiv:1209.5538}},
  \href {http://dx.doi.org/10.1016/j.nuclphysb.2012.11.018}
  {\path{doi:10.1016/j.nuclphysb.2012.11.018}}.

\bibitem{Ghezzi:2015vva}
M.~Ghezzi, R.~Gomez-Ambrosio, G.~Passarino, S.~Uccirati, {NLO Higgs effective
  field theory and kappa-framework}, JHEP 07 (2015) 175.
\newblock \href {http://arxiv.org/abs/1505.03706} {\path{arXiv:1505.03706}},
  \href {http://dx.doi.org/10.1007/JHEP07(2015)175}
  {\path{doi:10.1007/JHEP07(2015)175}}.

\bibitem{NLOnote}
G.~Passarino, M.~Trott, {{The Standard Model Effective Field Theory and Next to
  Leading Order}}, {LHCHXSWG-DRAFT-INT-2016-005,
  https://cds.cern.ch/record/$2138031$} (2016).

\bibitem{Hartmann:2015oia}
C.~Hartmann, M.~Trott, {On one-loop corrections in the standard model effective
  field theory; the $\Gamma(\PH \rightarrow \PGg \, \PGg)$ case}, JHEP 07
  (2015) 151.
\newblock \href {http://arxiv.org/abs/1505.02646} {\path{arXiv:1505.02646}},
  \href {http://dx.doi.org/10.1007/JHEP07(2015)151}
  {\path{doi:10.1007/JHEP07(2015)151}}.

\bibitem{Hartmann:2015aia}
C.~Hartmann, M.~Trott, {Higgs Decay to Two Photons at One Loop in the Standard
  Model Effective Field Theory}, Phys. Rev. Lett. 115~(19) (2015) 191801.
\newblock \href {http://arxiv.org/abs/1507.03568} {\path{arXiv:1507.03568}},
  \href {http://dx.doi.org/10.1103/PhysRevLett.115.191801}
  {\path{doi:10.1103/PhysRevLett.115.191801}}.

\bibitem{Grzadkowski:2010es}
B.~Grzadkowski, M.~Iskrzynski, M.~Misiak, J.~Rosiek, {Dimension-Six Terms in
  the Standard Model Lagrangian}, JHEP 1010 (2010) 085.
\newblock \href {http://arxiv.org/abs/1008.4884} {\path{arXiv:1008.4884}},
  \href {http://dx.doi.org/10.1007/JHEP10(2010)085}
  {\path{doi:10.1007/JHEP10(2010)085}}.

\bibitem{Boggia:2016asg}
M.~Boggia, R.~Gomez-Ambrosio, G.~Passarino, {Low energy behaviour of standard
  model extensions}, JHEP 05 (2016) 162.
\newblock \href {http://arxiv.org/abs/1603.03660} {\path{arXiv:1603.03660}},
  \href {http://dx.doi.org/10.1007/JHEP05(2016)162}
  {\path{doi:10.1007/JHEP05(2016)162}}.

\bibitem{Gavela:2016bzc}
B.~M. Gavela, E.~E. Jenkins, A.~V. Manohar, L.~Merlo, {Analysis of General
  Power Counting Rules in Effective Field Theory}\href
  {http://arxiv.org/abs/1601.07551} {\path{arXiv:1601.07551}}.

\bibitem{Kallosh:1972ap}
R.~E. Kallosh, I.~V. Tyutin, {The Equivalence theorem and gauge invariance in
  renormalizable theories}, Yad. Fiz. 17 (1973) 190--209, [Sov. J. Nucl.
  Phys.17,98(1973)].

\bibitem{Arzt:1993gz}
C.~Arzt, {Reduced effective Lagrangians}, Phys. Lett. B342 (1995) 189--195.
\newblock \href {http://arxiv.org/abs/hep-ph/9304230}
  {\path{arXiv:hep-ph/9304230}}, \href
  {http://dx.doi.org/10.1016/0370-2693(94)01419-D}
  {\path{doi:10.1016/0370-2693(94)01419-D}}.

\bibitem{David:2015waa}
A.~David, G.~Passarino, {Through precision straits to next standard model
  heights}, Rev. Phys. 1 (2016) 13--28.
\newblock \href {http://arxiv.org/abs/1510.00414} {\path{arXiv:1510.00414}},
  \href {http://dx.doi.org/10.1016/j.revip.2016.01.001}
  {\path{doi:10.1016/j.revip.2016.01.001}}.

\bibitem{D'Hoker:1984ph}
E.~D'Hoker, E.~Farhi, {Decoupling a Fermion Whose Mass Is Generated by a Yukawa
  Coupling: The General Case}, Nucl. Phys. B248 (1984) 59--76.
\newblock \href {http://dx.doi.org/10.1016/0550-3213(84)90586-8}
  {\path{doi:10.1016/0550-3213(84)90586-8}}.

\bibitem{Wess:1971yu}
J.~Wess, B.~Zumino, {Consequences of anomalous Ward identities}, Phys. Lett.
  B37 (1971) 95--97.
\newblock \href {http://dx.doi.org/10.1016/0370-2693(71)90582-X}
  {\path{doi:10.1016/0370-2693(71)90582-X}}.

\end{thebibliography}

\end{document}